\documentclass[12pt,a4paper]{article}
\usepackage[pdfstartview=FitH,colorlinks=true,linkcolor=blue,anchorcolor=red,citecolor=magenta,urlcolor=blue]{hyperref}
\usepackage[english]{babel}
\usepackage[utf8]{inputenc}
\usepackage{amsmath,amssymb,titling,authblk}
\usepackage{slashed}
\usepackage{amsmath}
\usepackage{amscd}
\usepackage[normalem]{ulem}
\usepackage{appendix}
\usepackage{bbold}
\usepackage{bm}
\usepackage{pdflscape}
\usepackage{yfonts}
\usepackage{amscd}
\usepackage{epsfig}
 \usepackage{microtype}
 \usepackage{upgreek}
 \usepackage{comment}

\usepackage{cite}

\allowdisplaybreaks[4]

\usepackage {color}
\usepackage{bbold}
\usepackage{braket}
\usepackage{bbm}

\advance\hoffset by -1.1truecm
\advance\voffset by -1.0truecm
\newcommand{\be}{\begin{equation}}
\newcommand{\ee}{\end{equation}}
\newcommand{\bea}{\begin{eqnarray}}
\newcommand{\eea}{\end{eqnarray}}

\newcommand{\del}{\partial}
\newcommand{\dd}{\mathrm d}
\newcommand{\ii}{\mathrm i}
\newcommand{\e}{\mathrm e}
\newcommand{\Tr}[1]{\:{\rm Tr}\,#1}
\usepackage{upgreek}
\newcommand{\sx}{ \mathrm x}
\newcommand{\ketbra}[2]{\ket{#1}\!\!\bra{#2}}

\newcommand{\la}{\langle}
\newcommand{\nn}{\nonumber\\}

\newcommand{\ba}{\begin{array}}
\newcommand{\ea}{\end{array}}
\newcommand{\1}{{\bf 1}}

\newcommand{\RR}{\mathbb{R}}

\newcommand{\RF}{\mathfrak{R}}
\newcommand{\rf}{\scriptscriptstyle\mathfrak{R}}

\newcommand{\sS}{\scriptscriptstyle S}
\newcommand{\srf}{\scriptscriptstyle{S\cup\mathfrak{R}}}

\newcommand{\HS}{\mathcal H_{\scriptscriptstyle S}}

\newcommand{\tlc}{\triangleleft}

\def \Hi {{\cal H}}

\newcommand{\rhogru}{{\uprho_{\scriptsize{\mathcal R}}}}
\newcommand{\rhoSR}{{\rho_{\srf}}}
\newcommand{\rhocarr}{{\rho_{\sS}}}

\usepackage{mathrsfs}
\usepackage{authblk}

\numberwithin{equation}{section}
\usepackage{float}
\restylefloat{figure}
\newcounter{appendice}

\usepackage[T1]{fontenc}
\usepackage{lmodern}

\begin{document}

\setlength{\droptitle}{-6pc}

\title{Mixed states for   
reference frame transformations}

\renewcommand\Affilfont{\itshape}
\setlength{\affilsep}{1.5em}

\author[1,3]{Gaetano Fiore\thanks{gaetano.fiore@na.infn.it, gaetano.fiore@unina.it}}
\author[2,3]{Fedele Lizzi\thanks{fedele.lizzi@na.infn.it, fedele.lizzi@unina.it}}
\affil[1]{Dipartimento di Matematica ``Renato Caccioppoli'', Universit\`{a} di Napoli {\sl Federico~II}, Napoli, Italy}
\affil[2]{Dipartimento di Fisica ``Ettore Pancini'', Universit\`{a} di Napoli {\sl Federico~II}, Napoli, Italy}
\affil[3]{INFN, Sezione di Napoli, Italy}

\date{}

\maketitle

\vspace{-2cm}

\begin{abstract}\noindent
We discuss the concept of  transformations among reference frames (classical or quantum). Usually transformations among classical reference frames have sharply defined parameters; geometrically they can be considered as \emph{pure states}  in the parameters' space, and they form a group. It is however possible that the distributions in the parameters' space are \emph{mixed states}; such states form a semigroup. Similarly, transformations among quantum reference frames can be either pure or mixed. 
{This gives rise to interesting consequences: the state of a system can be pure with respect to a reference frame and mixed with respect to another; we concretely discuss this in the framework of Galilei transformations in 1+1 dimensions. In particular, if the state of a reference frame with respect to another frame is thermal at some temperature, a quantum particle in  the pure (improper) rest state with respect to the first frame will appear in a thermal state with a
related nonzero temperature with respect to the other. 
This can also be discussed in relation to the time/energy uncertainty relation.}
 \end{abstract}

\newpage

\section{Introduction \label{intro}}

Descriptions of dynamical systems $\mathsf S$ require always a reference frame (RF), although this is often implicit. 
A RF $\RF$ is usually built using one or more macroscopic bodies  
with a huge number of atoms. Then,  typically, ``collective observable'' like the position 
of the center of mass,  or its total momentum, are not significantly affected by the observation of 
others systems, in particular of 
a system  of ``collective observables'' of another RF, $\RF'$. This is why  RFs are usually idealised as classical. 
But the ultimate quantum nature of these bodies will spoil their  classical (i.e.\  idealised)  properties via uncertainty relations and other quantum effects; this will be particularly manifest if the number of atoms is not so large, but also in other cases where coherence plays a role. This leads to the introduction of \emph{Quantum Reference Frames}.
The concept of Quantum Reference Frames (QRFs) was first proposed in~\cite{Aharonov:1967zza,Aharonov:1984zz}. The idea had been around for some time, and more
recently it has been argued that  QRFs may actually be a blessing, more than a source of complications, because the use of spacetime observables relative to QRFs could heal QFT divergences. In fact, Ref.~\cite{Jensen:2023ypr,Witten:2024ypr,Fewster2024,DeVuyst:2024pop,DeVuyst:2024uvd}, building on ideas and results of~\cite{Chandrasekaran:2022cip}, propose frameworks for local measurements of Quantum Fields on a symmetric background with respect to (w.r.t.) \  a QRF: under suitable assumptions the algebra of (relative) observables is a type II (rather than type III$_1$) factor, which entails the existence of a semifinite, or even finite, (instead of an infinite) trace, which allows, among other things, computing entropy.

Most of the recent papers dealing with QRFs use a ``Relational'' approach to Quantum Mechanics~\cite{Rovelli:1995fv,Loveridge:2018,Hoehn:2021PRD,Hoehn:2021FiP}; this means that a generic quantum system  is not in an unique ``absolute'' state, but in one state relative to each (quantum or classical) RF.
As a consequence, a composite system can be in an entangled state w.r.t.\ a QRF $\RF$, and at the same time in a factorized state with respect to another QRF $\RF'$. The field of quantum reference frames is currently very active; an incomplete list of references is:~\cite{Giacomini2019, Hamette2020, Vanrietvelde2020, Chen:2022wro, GiacominiBrukner:2020EEP, Giacomini2021, Amelino-Camelia:2022dsj, Hohn2022, Hamette2023, ApadulaCastro-RuizBrukner:2024}.
Moreover it appears that QRF's may find applications also at the level of table-top experiments, see for example~\cite{Tino2021Testing}.

Since observers, classical or quantum, use always a particular reference frame, and a change of observer implies a change of RF, the two concepts are closely related, especially in the quantum case. The difference between the two concepts is conceptually subtle, but in several cases, and for the scope of this paper, they can be treated as synonyms. Here we note that noncommutative spacetime requires the presence of quantum observers~\cite{FioreWess:2007,Fiore:2010,Lizzi:2018qaf, Lizzi:2019wto, Lizzi:2022hcq, Relancio:2024axb}. 

In the attempt to develop a consistent theory of Quantum Reference Frames - and their transformations - it is possible to proceed with a bottom-up approach, i.e.\ determining properties of QRFs starting from the quantum properties of their microscopic constituents, or a top-down approach, i.e.\ studying which classical properties of RFs are compatible with the quantum nature of RFs, or how they can be generalised. Whichever approach is used, the main theme of this paper is that not only the physical system to be observed, or the reference frames should be considered quantum objects, but also \emph{transformations} have to be considered as quantum\footnote{This means that the parameters of a transformation from a RF to another should not be considered as numbers, but as operators acting on a suitable Hilbert space.}, 
and as such may be in a pure or  mixed (i.e.\ nonpure) state\footnote{We use the terms mixed and nonpure states interchangeably.}. This is true also for \emph{classical} reference frames, which we consider here. In a related paper we will consider quantum reference frame transformations in a quantum (i.e.\ noncommutative) spacetime~\cite{FioreLizziinpreparation}.

Although the results are general, and we will consider Galilei transformations in Sect.~\ref{se:Galilei},  for most of this paper we will illustrate this with the simplest Lie group, that of translations. Even for this simple case we find dramatic conceptual consequences. In particular, even if the state of $\mathsf S$ is pure  w.r.t.\  the classical RF $\RF$, a nonpure transformation will transform it into a  mixed state. This means that if the state of (the collective degrees of freedom of) $\RF$ w.r.t.\  the classical RF $\RF'$ is mixed, then the state of {\sf S} w.r.t.\  $\RF'$ - inferred from the observation of $\RF$ and exchange of information with the latter - will be mixed. {In particular, if the state of $\RF$ w.r.t.\  $\RF'$ is thermal at some temperature $T_{\rf}$, a quantum particle in  the pure (improper) rest state w.r.t.\  $\RF$  will appear  w.r.t.\  $\RF'$ in a thermal state with  temperature $T'=T_{\rf} \tfrac mM$, where $m,M$ are the masses of the particle and of $\RF$ respectively. Considering either reference frame as a quantum object we sketch a possible connection between the time/energy uncertainty and the thermal states, thus giving a physical conceptual connection between time and temperature, without the necessity to introduce the complex plane.}

We begin in Sect.~\ref{RFTransf} with a description of the needed machinery to determine reference frame transformations, which is further described in a geometric way in Sect.~\ref{geom}. These introductory sections cover well known material, and serve mainly to fix notations. The core of the paper is Sect.~\ref{pureandmixed} where the concet of mixed transformation is introduced, and illustrated with examples. This section discusses the role of pure and mixed states as well. 
{Sect.~\ref{se:Galilei} discusses Galilei transformations in 1+1 dimensions, in particular the smeared changes of reference frames, and a possible connection between the time/energy uncertainty and the thermal states. The final section collects some final remarks.}

\section{Reference Frame Transformations \label{RFTransf}}

Changes of classical  reference frames $g:\RF\mapsto\RF'$ in
space(time) make up a  Lie group $G$: 
\begin{itemize}
\item the product $gg'$ is the composition of  $g,g'$; 
\item the unit is $\1:\RF\mapsto\RF$; 
\item the inverse of \ $g:\RF\mapsto\RF'$ \ is \ $g^{-1}:\RF'\mapsto\RF$. 
\end{itemize}
Any group element ${g}$ sharply specifies a transformation of RF, i.e., how $\RF$ is located and moves 
w.r.t.\ to $\RF'$. Denote by $\sx,\sx'$ the $n$-ples of spacetime coordinates used by the two different observers to identify a generic event
in their respective reference frames $\RF,\RF'$; ${g}$ determines the 1-to-1 map $\sx\mapsto \sx'$ (defined on all possible events). The latter induces  by pull-back a 1-to-1  map, denoted as a \emph{passive  transformation}, between the dynamical variables 
(vectors - like a particle momentum or angular momentum -, particle worldlines, fields, states, associated wavefunctions,\ldots) used by $\RF,\RF'$ to describe a generic physical system {\sf S}; \ for instance,  the map 
\be
\gamma({g}):\varphi\mapsto\varphi'
\label{PassiveTransf}
\ee
for a real scalar field $\varphi$ is determined imposing the relation \ $\varphi(\sx)=\varphi'[\sx'(\sx)]
$. The change in the coordinates' dependence is due to the change of the RF, not to a change of the observed system $\mathsf S$; 
this is sometimes alternatively referred to as
{\it bodily identity}  of the two physical descriptions \cite{Barut:1959gmp}.

If $g:\RF\mapsto\RF'$ is a change of inertial RFs
on flat spacetime, (\ref{PassiveTransf}) reduces to 
\bea
\ba{l}
\gamma({g}):\varphi\mapsto \varphi'
\equiv \varphi \tlc {g}, \\[4pt]
\varphi'(\sx')\equiv\varphi\big(\sx'{g}^{-1}\big)
\ea
\label{Poincare_phiphi'}
\eea 
The four-component vectors $\sx,\sx'$ are  related by
\be
\sx'_\mu=( \sx{g})_\mu\equiv\sx_\nu\Lambda^\nu_\mu + a_\mu,
\label{Poincare_xx'}
\ee
where  ${g}\!\equiv\!(\Lambda,\!a)\!\in\! G\equiv$ Poincaré or Galilei group. Here $a\in \RR^4$ is the spacetime translation vector, while $\Lambda\in $ is the matrix representing the Lorentz (or homogeneous Galilei) transformation.

The maps $\gamma$ apply with the same form whether ${\sf S}$ is classical or quantum;
e.g., whether the fields $\varphi$ are  $c$-number-valued or operator-valued, and, in the former case, whether  $\varphi$ is a classical field (like the atmospheric pressure) or a wavefunction, or a (scalar) quantum field.

Enforcing the maps $\gamma$ requires that $\RF'$ has: 
\begin{itemize}
\item[i)] full information about the description of {\sf S} by $\RF$;
\item[ii)] sharply determined $g$, i.e.\ how $\RF$ is located and moves with respect to
(wrt) $\RF'$.
\end{itemize}

On (flat) spacetime  
the laws of physics appear exactly the same to $\RF,\RF'$, due to their form-invariance under   (Poincar\'e) transformations. Therefore both $\RF,\RF'$ can prepare a copy
of the same experiment in such a way that the two involved physical system ${\sf S},{\sf S}'$
resp.\ appear to $\RF,\RF'$ exactly in the same way; in particular $\RF,\RF'$  will identify a pair
$(E,E')$ of corresponding events  in the two experiments through the same values of their 
coordinates: $\sx_{\scriptscriptstyle E}\!=\!\sx'_{\scriptscriptstyle E'}$. 
This is sometimes referred to as
{\it subjective identity}  of the two physical descriptions \cite{Barut:1959gmp}. 
The correspondence 
between the dynamical variables (states, observables, etc.)
used by $\RF$ to describe ${\sf S},{\sf S}'$,
  is an \emph{active Poincar\'e transformation}\footnote{In the literature active transformations are often introduced thinking of ${\sf S}'$ as   ${\sf S}$ {\it moved} so as to reach the same position and velocity which would have  ${\sf S}'$. Since this would involve  (temporary) accelerations/decelerations of ${\sf S}$ w.r.t.\  $\RF$, which would modify its dynamics, we prefer to avoid such a way of defining active transformations.}. 
  Again, it depends 
only on the element of the group relating $\RF,\RF'$. 
In a quantum theory this correspondence
is a strongly continuous unitary representation $U$ of
$G$  on the Hilbert space $\mathcal H$ of the system: 
$\Psi\in\HS\mapsto \Psi^g=U(g)\Psi\in\HS$ for the pure states,
 $\alpha\mapsto \alpha^g=U\!(g)\,\alpha\, U^{\dagger}\!(g)$ 
for the observables  on $\HS$. 
By the above definition, the corresponding wavefunction $\psi^g$ w.r.t.\  $\RF$ will equal $\psi'$ (up to a global phase), i.e.\  $\psi^g(\sx)=\psi'(\sx)=\psi(\sx g^{-1})$.

Since in our conventions of eq.~(\ref{Poincare_xx'}) a Lorentz transformation acts on the coordinates by matrix multiplication from the {\it right}, both maps $g\mapsto \gamma(g)$,  $g\mapsto U(g)$, are group antihomomorphisms, 
\be
[\gamma(gg')\varphi](\sx)=\varphi\big[\sx(gg')^{-1}\big]=\varphi\big(\sx g'^{-1}g^{-1}\big)
=[\gamma(g)\varphi]\big(\sx g'^{-1}\big)=[\gamma(g')\gamma(g)\varphi]\big(\sx\big)
\ee
Therefore $\gamma(gg')\!=\!\gamma(g')\gamma(g)$, 
$U(gg')\!=\!U(g')U(g)$. In other words, they are right $G$-actions\footnote{One could also choose conventions with right and left interchanged.}. 

As we see, one can transform the dynamical variables of a system either 
by a passive transformation, or an active transformation, or even, more generally, any mixture of
the two. These will be respectively parametrised by elements $(g,\1),(\1,g'),(g,g')\in G\times G$.

All the transformations described here assume a full knowledge of the parameters of the transformation. In other words, no matter how the states are blurred, it is always assumed that the parameters of the transformation are known with absolute precision. We will question this assumption in the following.

\section{The Geometry of classical RF Transformations \label{geom}}

The sharp RF transformations we described earlier form a Lie group $G$, meaning that each possible transformation is a point of a group manifold. Also transformations among  QRFs may form a group $G$~\cite{Ballesteros:2020lgl};  indeed ref.~\cite{Ballesteros:2025ypr} shows that  the transformations among nonrelativistic QRFs form a generalized Galilei group. This is one of the motivations of our analysis in sect.~\ref{se:Galilei}.
The group, being also a topological space, as any manifold, can be described by the algebra of continuous functions defined on it. For the noncompact case it is required also that functions vanish on the boundary; in this case the algebra will not contain the unity. These functions make up a  $C^*$-algebra $C(G)$, and
all topological informations of the space are contained in this $C^*$-algebra\footnote{For a mathematical treatment see for example~\cite{FellDoran1988}, a physicist's friendly introduction is in~\cite{Lizzi:2018dah}.}.

The points of $G$, considered as a manifold, can be reconstructed as \emph{pure states}  from $C(G)$, or in the presence of a boundary, from a  subalgebra $\mathcal A$  of $C(G)$ consisting of test functions (e.g. smooth functions on $G$ rapidly going to zero at the boundary).
A state for an algebra $\mathcal A$ is a linear functional $\rhogru:f\in\mathcal A \to \mathbb C$ with the properties:

\begin{enumerate} 

\item Positivity: $\rhogru(f^\dag f)\geq 0$.

\item Unit norm: $\|\rhogru\|=\sup_{\|f\|\leq 1} |\rhogru(f)|=1$. 

\item If the algebra has a unity $\mathbb 1$ (compact groups have it) then $\rhogru(\mathbb 1)=1$.

\end{enumerate}

States can be combined, they form a convex space: given two states $\rhogru_1,\rhogru_2$, and a number  $0<\lambda<1$ the sum $\lambda \rhogru_1+(1-\lambda)\rhogru_2$ is still a state.
Some states cannot be expressed as such convex sum for any choices of $\rhogru_i$ and $\lambda$, they form
the boundary of the set and are called \emph{pure states}.

A generic (not necessarily pure) \emph{state} functional  $\rhogru$ is  identified by a probability density  (i.e.\  a positive definite normalised function or distribution) $\widetilde{\rhogru}$, via\footnote{Sometimes in the literature, with an abuse of notation, the functional and the corresponding distribution are denoted by the same symbol; we avoid this confusion in the following.}:
\be
\rhogru(f)=\int\dd a f(a) \widetilde{\rhogru}(a). \label{genericrho}
\ee
A particular class of states is given by the Dirac's $\delta$ distributions: if $\rhogru=\delta_{a_0}$, then
$\widetilde{\rhogru}(a)=\delta(a-a_0)$.

As we said, a \emph{pure state} is a state that cannot be expressed as a convex sum of two other state. In the case above a generic positive normalised function can be expressed as a sum of two other functions, while the state identified by the Dirac's $\delta$ cannot.  Points are identified with the pure states, and the topology can be reconstructed by convergence.
A set of points/states $\delta_n$ converges to $\delta$ if
\be
\lim_n \delta_n=\delta \ \mbox{if}\ \forall f\in\mathcal A\ \mbox{then}\ \lim_n\delta_n(f)=\delta(f) \label{converg}
\ee

Let us look at the example of translations. As a group they make up the real line $\mathbb R$.
In principle we should start with a vector space over $\mathbb C$, on which a sum and a product among the element is defined, and a norm. The vector space however has infinite dimensions. Therefore we work in reverse and show that indeed the pure states are the points for the algebra of continuous functions vanishing at infinity. This section's main purpose is to fix notations, and we therefore sacrifice mathematical rigour and completeness.

Indicate the functions on the group as $f(a)$, the norm is the \emph{sup norm}:
\be
\|f\|=\sup_a|f(a)|
\ee
A state that can be identified by a positive normalised function $\widetilde\rhogru(a)$ can be the sum of two other states since it is always possible to find a pair of normalised positive functions $\widetilde\rhogru_i$ with $\lambda\widetilde\rhogru_1+(1-\lambda)\widetilde\rhogru_2=\widetilde\rhogru$. This procedure is not possible for the $\delta$ distributions, and we found that the set of pure states correspond to the points of the line. It is easy to verify that the topology defined by~\eqref{converg} is the usual one for the real line.
 
A group $G$ is not a mere topological space, there is a relation among the points. Namely: given two points the product associates a third one, there is a special point, the identity, and associated to every point there is another point, called the inverse, with all known properties. The algebra of functions on the group $C(G)$ is a Hopf algebra, which in addition to the property of algebras (internal sum and product, product by a complex number) has also three more structures. 
\begin{itemize}
\item The coproduct $\Delta: C(G)\to C(G)\to C(G)\otimes C(G)$. At the level of the algebra the product among points is dually reflected in the coproduct in the algebra:
\be
\Delta f=\sum_i f_i\otimes f_i ,\qquad\quad \Delta f (a_1,a_2)= f(a_1+a_2) \label{coprodtrans}
\ee  
where of course in the generic case the + must be substituted by the group product.

\item The counit $\epsilon: C(G)\to \mathbb C$ identifies the point corresponding to the identity: $\epsilon(f)=f(0)$. Where in the general case 0, the identity for the translation, shoud be substituted by the notation for the identity of the group.

\item The antipode $S:C(G)\to C(G)$ encodes the inverse $(Sf)(a)=f(-a)$. Again for a general group $-a$ should be $g^{-1}$.

\end{itemize}

\section{Pure and Mixed Translations in One Dimension \label{pureandmixed}}

Let us first consider translations in one dimension. Translations in this case are points of the real line $\mathbb R$ with the group structure given by the addition. We will indicate the points (and the coordinate functions) of the configuration space of a quantum particle {\sf S} by the letters like $\sx,\sx_0,\ldots$ A wave function, i.e.\ a pure state, will therefore be indicated by $\Uppsi(\sx)$. When we act with the group of translations we will instead use the letters $a,a_0\ldots$. A pure state on the group is identified by a value $a_0$, we will represent the element of the group acting on functions of $\sx$ as\footnote{Remember that in the example the identity is given by $a=0$ and that ${\frac\del{\del \sx}}^\dag=-\frac\del{\del \sx}$.}: 
\be
\pi(a_0)\Uppsi(\sx)=U_{a_0}\Uppsi(\sx)=\e^{a_0\del_\sx}\Uppsi(\sx)=\Uppsi(\sx+a_0)
\ee
Interpreting $\Uppsi(\sx)=\left\langle\sx|\Uppsi\right\rangle$ as a ``wave function'', i.e.\  $\Uppsi(\sx)\in L^2(\mathbb R)$, we have given the representation of pure states of the group over the  pure states of {\sf S}. We will call the Hilbert space $\Hi\simeq L^2(\mathbb R)$  of the $\Uppsi$'s the \emph{carrier space}.

The wave functions are not the only states of {\sf S}, there are also mixed states. We can easily see the transformation of a mixed state of the carrier space. To distinguish the two density matrices we use the notation\footnote{Do not confuse the states $\rhogru$ on the group, with the states $\rhocarr$ on the carrier space.} $\rhocarr$ for these other density matrices. Then
\be
\pi(a_0)(\rhocarr)=\e^{a_0^\mu\frac\del{\del\sx^\mu}}\rhocarr\e^{-a_0^\mu\frac\del{\del\sx^\mu}}=U(a_0)\rhocarr U(a_0)^\dag \label{puregrouptranf}
\ee
which for  $\rhocarr=\ketbra\Uppsi\Uppsi$ reduces to the previous case, and of course $\langle\sx \ketbra\Uppsi\Uppsi\sx\rangle=|\Uppsi(\sx)|^2$. Being unitary the representation transforms pure states in pure states. All this is well known.

It is possible to interpret this translation as the action of a pure state on the group (a single point on the group manifold) on the set of states of $\sf S$, via~\eqref{puregrouptranf}. Consider now nonpure states for the algebra of functions \emph{on the group}. This means that we are not considering a single point on the group manifold (an element of the group), but a density probability on the group manifold. In other words, we are not translating by a definite amount, but rather we have a certain probability to have a particular translation.

The possibility of having quantized frame transformations has being considered recently in~\cite{DEsposito:2024wru}. There are points of contacts with this work, but in this paper we consider changes of classical reference frames and not quantum groups, which will be treated in~\cite{FioreLizziinpreparation}.

Let us find an action of these nonpure states on the carrier space, some sort of ``$\pi(\rhogru)$''. As we will see, there is no group structure on the space of states, and therefore we cannot find a representation as unitary operators. We will nevertheless define an action of the space of state on the carrier space and use the same symbol $\pi$.

Start with the simplest nonpure  state 
\be
\widetilde{\rhogru}_{a_1,a_2}(a)=\frac12\delta(a-a_1)+\frac12\delta(a-a_2) \label{tilderho12}
\ee
(built as a convex sum of two pure  states), so that 
\be
\rhogru_{a_1,a_2}(f)=\frac12f(a_1)+\frac12f(a_2) \label{halfhalftrans}
\ee
this state just calculates the average of the function $f$ at two points. Intuitively this state is a weighted sum of two translations, or in general two transformations. 

Consider now the action on the carrier space. We need to reproduce the weighted sum of~\eqref{halfhalftrans}, which translates into a weighted sum of two translation by the amounts $a_1$ and $a_2$.  Therefore the nonpure state $\rhogru$ will act on states of the carrier space as:
\be
\pi(\rhogru_{a_1,a_2})(\rhocarr)=\frac12U(a_1)\rhocarr U(a_1)^\dag+\frac12U(a_2)\rhocarr U(a_2)^\dag \label{uprhohalfhal}
\ee
For the case $\rhocarr=\ketbra\uppsi\uppsi$ we have, in the position representation
\be
\bra \sx \pi(\rhogru_{a_1,a_2})\rhocarr\ket \sx=\Tr\big(\ketbra \sx \sx \pi(\rhogru_{a_1,a_2})\rhocarr\big)=\frac12\left|\uppsi(\sx+a_1)\right|^2+\frac12\left|\uppsi(\sx+a_2)\right|^2.
\ee

The r.h.s.\  of~\eqref{uprhohalfhal} is still a density matrix, but even if we started with a pure state of $\sf S$
w.r.t.\ $\RF$, we end up with a mixed state of $\sf S$
w.r.t.\ $\RF'$! There is no contradiction in this, because the whole state we started with (the state of $\RF\cup{\sf S}$ w.r.t.\ $\RF'$) was mixed.  

We can easily generalise this to a generic $\rhogru(a)$ as in~\eqref{genericrho}, in this case we have:
\be
\pi(\rhogru)\rhocarr=\int \dd a \widetilde{\rhogru}(a) U(a)\rhocarr U(a)^\dag
\ee
and for pure states $\rhocarr=\ketbra\uppsi\uppsi$ 
\be 
\Tr\big(\ketbra \sx \sx \pi(\rhogru)\rhocarr\big)= \int \dd a\, \widetilde{\rhogru}(a)\, |\uppsi(\sx+a)|^2. \label{trasladensity}
\ee
{From the mathematical point of view, it is possible to see such a RF transformation as an averaging over the group (G-twirling, see for example~\cite{TolouiGour}) with a nontrivial\footnote{In the sense that is neither
concentrated in a single point, nor uniform.}  probability measure.}

Let us now look for product, identity and inverse. The group is abelian, therefore certain aspects may be simplified. 
 Given two generic states  $\rhogru_1,\rhogru_2$, we define the product state, using the coproduct defined in~\eqref{coprodtrans}, as 
\be
(\rhogru_1\rhogru_2)(f)=\int\dd a\dd a' \widetilde{\rhogru}_1(a)\widetilde{\rhogru}_2(a') \Delta f(a,a')=\int\dd a\dd a' \widetilde{\rhogru}_1(a)\widetilde{\rhogru}_2(a') f(a+a'); \label{prodvecc}
\ee
when $\widetilde{\rhogru}_i(a)=\delta(a-a_i)$, i.e.\  the states are pure, we have 
\be
(\rhogru_1\rhogru_2)(f)=f(a_1+a_2)
\ee
as wished. 
 It is straightforward  to check that (\ref{prodvecc}) can be written in the form (\ref{genericrho}) with a density function $\widetilde{\rhogru}(a)$ given by the convolution
\be
\widetilde{(\rhogru_1\rhogru_2)}(a)=\int\dd b \,\, \widetilde{\rhogru}_1(b) \,\widetilde{\rhogru}_2(a-b)
\label{prodGenGroupStates}
\ee
even when $\rhogru_1,\rhogru_2$ are not pure; the purity of states plays no role in this definition. 
Here in the general case \ $a-b$ \ is to be understood as the group product of $a$ by the inverse of $b$. 

The identity functional  has the density 
$\widetilde{\rhogru}(a)=\delta(a)$ as before; it is related to the counit defined earlier.  It remains to verify the existence of the inverse to give a full group structure. At the level of the algebra the inverse is given by the antipode $S(f)$, which in this simple case is simply $[S(f)](a)=f(-a)$. To the pure state with density $\delta(a-a_1)$ there corresponds the inverse  pure state with density $\delta(-a-a_1)$. This does not work for the mixed states!  Take the simplest example $\rhogru_{a_1,a_2}$. If we act first with this state, and then with the state $\rhogru_{-a_1,-a_2}$, the result is
\begin{align}
\rhogru_{-a_1,-a_2}\left(\rhogru_{a_1,a_2}(\rhocarr)\right)&=\rhogru_{-a_1,-a_2}\left(\frac12
U(a_1)\rhocarr U(a_1)^\dag+\frac12U(a_2)\rhocarr U(a_2)^\dag\right)\nonumber\\
&=\frac12\rhocarr+\frac14U(a_1-a_2)\rhocarr U(a_1-a_2)^\dag+\frac14U(a_2-a_1)\rhocarr U(a_2-a_1)^\dag
\end{align}
{In principle there may be another nonpure state which acts as inverse, but it is possibile to prove that there are elements which do not have inverse. 
Take for example the half translation defined in~\eqref{tilderho12}. Assume that there is a $\widetilde{\rhogru}_{a_1,a_2}^{-1}(a)$ corresponding to the inverse. Then, inserting it in~\eqref{prodGenGroupStates}, we should find the density function corresponding to the identity transformation, namely the $\delta$. Explicitly, we should have
\begin{align}
\delta(a)&=
\int\dd b \,\, \widetilde{\rhogru}_{a_1,a_2}^{-1}(b)\,\widetilde{\rhogru}_{a_1,a_2}(a-b)\nonumber\\
&=\frac12\int\dd b \,\, \widetilde{\rhogru}_{a_1,a_2}^{-1}(b)\,\delta(a-b-a_1)+
\frac12\int\dd b \,\, \widetilde{\rhogru}_{a_1,a_2}^{-1}(b)\,\delta(a-b-a_2)\nonumber\\
&=\frac12\left(\widetilde{\rhogru}_{a_1,a_2}^{-1}(a-a_1)+\widetilde{\rhogru}_{a_1,a_2}^{-1}(a-a_2)\right)
\end{align}
the function $\rhogru$ cannot be the identity, and therefore the two functions above are different, and they should define two different states for $a_1\neq a_2$. Since the $\delta$ defines a pure state it cannot be written as convex sum of two other pure states, and therefore we have shown that this element ha no inverse. This procedure can easily be generalised to show that all nonpure states on the group have no inverse. }

On the contrary, pure states have an inverse, given by the Hopf algebra structure via the antipode, so that the inverse of the pure state $\delta(a-a_1)$ is $\delta(a+a_1)$, since for translations the antipode of the element $a_1$ is $(-a_1)$.

We conclude that instead of a group we have a semigroup. 
In the dual algebraic setting, a semigroup amounts to a \emph{bialgebra}, i.e.\ an algebra endowed with compatible coproduct and counit. With respect to the Hopf algebra case, the antipode is missing. 

{As we already mentioned after Eq.~\eqref{trasladensity}, what we call smearing can be seen as an averaging over the group in the presence of a nontrivial measure. The measure theory on Lie group is discussed in the mathematics literature, see for example~\cite{Meyer} especially in the context of diffusion processes and generalizations of the central limit theorem, and the presence of semigroups in this case is known. Here we connect it to representations of states and the quantization of reference frames.}

\begin{figure}[ht]
    \centering
   \includegraphics[scale=0.9]{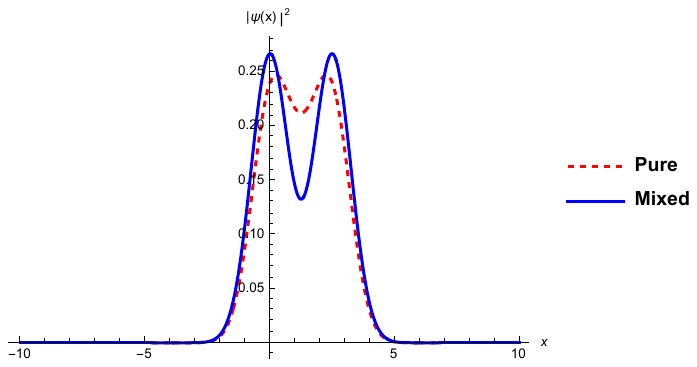}
    \caption{\textsl{
    The probability density for the pure and mixed states for  the sum of two Gaussians. The chosen parameters are $\upalpha=0.75, a_2=2.5$. }}
    \label{fig_a1a2}
\end{figure}

\begin{figure}[ht]
    \centering
   \includegraphics[scale=0.9]{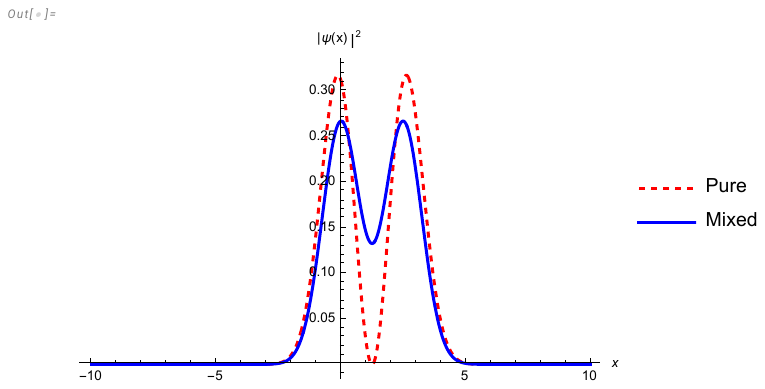}
    \caption{\textsl{
      The probability density for the pure and mixed states for the difference between two Gaussians. The chosen parameters are $\upalpha=0.75, a_2=2.5$. }}
    \label{fig_a1a2diff}
\end{figure}

Let us discuss an example in which we start from the pure gaussian state:
\be
\uppsi(\sx)=\frac1{\sqrt{\upalpha\sqrt{2\pi}}}\e^{-\frac{\sx^2}{4\upalpha^2}}
\ee
We consider first the case of $\widetilde\rhogru_{a_1,a_2}$ of~\eqref{tilderho12}, setting for simplicity $a_1=0$. The translated mixed state has the following probability density of finding the particle in $\sx$:
\be
\frac12\left|\uppsi(\sx)\right|^2+\frac12\left|\uppsi(\sx+a_2)\right|^2
=\frac{\e^{-\frac{\left(\sx-a_2\right){}^2}{2 \upalpha
   ^2}}+\e^{-\frac{\sx^2}{2
   \upalpha ^2}}}{2 \sqrt{2 \pi
   } \upalpha }
  \ee
This can be compared with the pure state case of a single wave function that is the sum or difference of two Gaussians centered in 0 and $a_2$ and the probability density
\be
\uppsi_{a_2}(\sx)=N\left(\e^{-\frac{\sx^2}{4\upalpha^2}}\pm\e^{-\frac{(\sx-a_2)^2}{4\upalpha^2}}\right)
\ee
with $N$ a new normalization constant necessary because the transformation is non unitary. Then (after some calculations) we have
\be
|\uppsi_{a_2}(\sx)|^2=\frac{\left(\e^{-\frac{\left(\sx-
   a_2\right)^2}{4 \upalpha
   ^2}}\pm \e^{-\frac{\sx^2}{4
   \alpha ^2}}\right)^2}{2
   \sqrt{2 \pi } \upalpha 
   \left(\e^{-\frac{a_2^2}{8
   \alpha ^2}}+1\right)}
\ee
The absence of the mixed terms in the mixed state has the effect to divide the two maxima in a sharper way. This can be seen in Fig.~\ref{fig_a1a2}.
The net effect is that the mixed state, which divides more the two peaks, is more ``classical'' than the pure  state obtained as a sum of Gaussians; for the difference, the converse is true as it can be seen in Fig.~\ref{fig_a1a2diff}. 

We now consider the case in which also 
the nonpure translation is a Gaussian: 
\be
\widetilde\rhogru(a)=\frac1{\sigma\sqrt{2\pi}}\e^{-\frac{(a-a_0)^2}{2\sigma^2}} \label{rhotiledea}
\ee
Using~\eqref{trasladensity} we find that for the new state
\be
\int \dd a\, \widetilde{\rhogru}(a)\, |\uppsi(\sx+a)|^2=\frac1{\sqrt{\sigma^2+\upalpha^2}\sqrt{2\pi}}\e^{-\frac{(\sx-a_0)^2}{2(\sigma^2+\upalpha^2)}}
\ee
This appears as a simple spreading of the wave packet, but this would be misleading, the state resulting from the nonpure translation is nonpure and cannot be described by a single function of $\sx$. One can consider the pure state obtained translating in a similar Gaussian way a Gaussian wave function. The state would be
\be
\uppsi_{\mbox{transl}}(\sx)=N\int \dd a\, \widetilde{\rhogru}(a)\uppsi(\sx+a)
\ee
 In this case we have (after some calculations)
\be
|\uppsi_{\mbox{transl}}(\sx)|^2=
\frac{\e^{-\frac{(\sx-a_0)^2}{\sigma ^2+2 \upalpha^2}}}
{\sqrt{\pi }   \sqrt{\sigma ^2+2\upalpha^2}}. \label{xtranslate}
\ee
Since $2(\sigma^2+\upalpha^2)>\sigma ^2+2\upalpha^2$ we see that the pure state is always more localised than the mixed one.

\section{Mixed states of RF (1+1)-dim Galilei transformations and thermal states\label{se:Galilei}}

The Galilei Lie algebra in 1+1 dimension is generated by the momentum $\hat p$ for space translations, the Hamiltonian $\hat H$  for time translations,  $\hat K$ for  boosts, together with a central element $\hat m$ (the mass). These have the Lie brackets
\be
[\hat p,\hat H]=0, \qquad [\hat K, \hat p]=\ii\hbar\hat m,  \qquad [\hat K, \hat H]=\ii\hbar \hat p, \qquad [\hat m,\:\cdot\:]=0.
\label{GalileiCR}
\ee
On the Hilbert space of a free quantum particle of mass $m$, space coordinate $\hat \sx  $ and momentum $\hat p$ the last three generators can be realized as the hermitean operators $\hat H=\hat p^2/2m$, $\hat K=\hat m\hat \sx  -t\hat p$, and $\hat m=m$, where of course $[\hat\sx,\hat p]=\ii\hbar$. These  satisfy the commutation relations~(\ref{GalileiCR}).

The unitary operator corresponding to a ``sharp'' RF change $g_v:\RF\mapsto \RF'$, where
the origin of $\RF$ has velocity $v$ with respect to  $\RF'$, is $\e^{\ii\frac v{\hbar}\hat K}$. Using 
the Baker-Campbell-Hausdorff formula, the expression can be rewritten as:
\be
\e^{\ii \frac v{\hbar}\hat K}=\e^{\ii \frac {mv}{\hbar}\hat \sx  }\e^{-\ii t\frac {v}{\hbar} \hat p}\e^{-\ii t\frac {mv^2}{2\hbar}}.
\ee 
Given any generalized pair of eigenstates $\ket{\sx  },\ket{p}$ resp.\ of $\hat \sx  ,\hat p$ w.r.t.\  $\RF$, $\hat \sx  \ket{\sx  }=\sx  \ket{\sx  }$, $\hat p\ket{p}=p\ket{p}$, we have $\la \sx  \!\ket{p}=\e^{-\ii p\sx  /\hbar}$, and
\bea
\bra{\sx  } \e^{\ii \frac v{\hbar}\hat K}\ket{p}&=&\bra{\sx  }\e^{\ii \frac {mv}{\hbar}\hat \sx  }\e^{-\ii t\frac {v}{\hbar} \hat p}\e^{-\ii t\frac {mv^2}{2\hbar}}\ket{p}=\la \sx  \!\ket{p} \e^{-\frac \ii{\hbar}(mv\sx  +tvp-tmv^2\!/2)}\nn
&=&\e^{-\frac \ii{\hbar}[(p+mv)\sx  +tvp-tmv^2\!/2]}
=\la \sx   \ket{p+mv} \e^{-\frac {\ii t}{\hbar}(vp-mv^2\!/2)}; \nonumber
\eea
one then finds
\be
\e^{\ii \frac v{\hbar}\hat K}\ket{p}=\ket{p+mv}\e^{-\frac {\ii t}{\hbar}(vp-mv^2\!/2)}.
\ee
If the state of $\RF$ w.r.t.\  $\RF'$ is a mixed state   $\rhogru$ described by a distribution
$\widetilde\rhogru(v)$, while that of  ${\sf S}$ w.r.t.\   $\RF$  is $\rhocarr=\ket{p}\!\bra{p}$, then the state of ${\sf S}\cup\RF$ will be \
$\rhoSR=\ket{p}\!\bra{p}\otimes \rhogru$, \ and the state \ $\rhocarr'$ \ of ${\sf S}$ w.r.t.\   $\RF'$ will be the mixed state
\be
 \rhocarr'=\mbox{tr}_{\mathcal H_{\rf}}\left(\e^{\ii \frac v{\hbar}\hat K}\rhoSR\,\e^{-\ii\frac v{\hbar}\hat K}\right)=  \int \!\dd v\, \widetilde\rhogru(v)\, \ket{p\!+\!mv}\!\bra{p\!+\!mv}.  \label{smearboost}
\ee
 In other words, a nonzero standard deviation\footnote{We use the same symbol $\Delta$ for both the uncertainty (or the standard deviation) and the coproduct. The difference should be clear from the context.} $\Delta v$ for the
distribution of the relative velocity $v$ of the reference frames, or equivalently $\Delta E_{\rf}$ for the
distribution of relative kinetic energy $E_{\rf}=\tfrac{Mv^2}2$ (here $M$ is the mass of $\RF$), transforms a pure state into a (proper) mixed state for the particle, hence leads also to a
nonzero uncertainty $\Delta p$ for the momentum of the particle. We note in addition that this occurs even though, as  in the present case, the pure state is improper (i.e.\ $\ket{p}$ is not normalizable and therefore does not belong to $\HS$; it is only the limit, in the distributional sense, of a sequence of proper states).

Let us apply this to the case of a quantum particle at rest with respect to the reference frame  $\RF$, i.e.\ its state with respect to  $\RF$ is described by the density matrix $\rhocarr_0=\ketbra{0}{0}$. The two reference frames need not be in relative motion with respect to each other, their relative \emph{average} velocity may be zero; in this case the transformation will be a smearing of the identical transformation. In particular, it is natural to consider the case that $\RF$ is in a thermal state with respect to the other reference frame $\RF'$, i.e. $\rhogru_{\beta_{\rf}}$ \ is the thermal state at some temperature $T_{\rf}$, which is described by the distribution
\be
\widetilde\rhogru_{\beta_{\rf}}(v)=\sqrt{\frac{M{\beta_{\rf}}}{2    \pi}} \e^{-\frac{Mv^2}{2}{\beta_{\rf}}}\,,  \qquad\quad {\beta_{\rf}}=\frac 1{K_BT_{\rf}}\,,
\label{Gausme}
\ee
where $K_B$ is Boltzmann constant. 
This is a thermal distribution  of relative velocities   - therefore also of relative momenta/energies - of  \emph{$\RF$ with respect to  $\RF'$}; since it has zero average velocity,  $\RF$ is in average at rest with respect to $\RF'$.
Correspondingly, $\Delta v= 1/\sqrt{M\beta_{\rf}}$, and $\Delta E_{\rf}= 1/{\beta_{\rf}}\sqrt{2}$. 
The parameter ${\beta_{\rf}}$  has the dimension of an inverse energy and controls the amount of smearing: a large ${\beta_{\rf}}$ describes a precise knowledge of the momentum, and since this is centred around zero momentum, a low average energy, and consequently a low temperature; on the other extreme when ${\beta_{\rf}}$ is small the system is at high temperature, with a poor knowledge of the energy.
Using~\eqref{Gausme} the state~\eqref{smearboost} becomes
\be
 \rhocarr'=  
\int \!\dd v\, \,\sqrt{\frac{M{\beta_{\rf}}}{2    \pi}} \e^{-\frac{Mv^2}{2 }{\beta_{\rf}}}
 \ket{mv}\!\bra{mv}=  \int \!\dd q\, 
\sqrt{\frac{\beta'}{2 m   \pi}} \e^{-\frac{q^2}{2m }\beta'}  \ketbra{q}{q}, \label{thermstate}
\ee
where we have defined $\beta'={\beta_{\rf}} M/m$ and the momentum $q=mv$.
The amount of mixing of the state~\eqref{thermstate} is   dictated by  parameter $\beta'$. While the state $ \rhocarr=\ketbra{0}{0}$ has temperature \ $T=0$ \ with respect to  $\RF$, we recognise that~\eqref{thermstate} is the  \emph{thermal state} of  $\sf S$  at temperature 
\be
T'=T_{\rf} \frac m{M}, 
\label{modifiedT}
\ee
wrt $\RF'$, because $\frac1{\sqrt{2\pi m K_BT' }} \e^{-\frac{q^2}{2mK_BT'}}$
is the distribution of momenta $q$ in a gas at temperature $T'$. In the limits  $T_{\rf}\to 0$,
or $M\to\infty$ (classical RFs), we find $T'\to 0$ and recover the pure state $ \rhocarr'=\ketbra{0}{0}$.  
If  instead $\rhogru$ is a thermal state with  temperature $T_{\rf}$ peaked around some nonzero average velocity $v_0$, i.e.
\be
\widetilde{\rhogru}_{\beta_{\rf},v_0}(v)= \sqrt{\tfrac{M}{2\pi K_BT_{\rf}}}\:
e^{-\frac{M(v-v_0)^2}{2K_BT_{\rf}}},
\ee
and $\rhocarr_0=\ketbra{p}{p}$ with generic $p\in\RR$,
then  $\rhocarr'$ will be in a thermal state with  temperature $T'$, but with momentum peaked around  $p'\equiv p+mv_0$:
\bea
\rhocarr'&=&  \int \!\dd v\, \,\sqrt{\tfrac{M{\beta_{\rf}}}{2    \pi}} \e^{-\frac{Mv^2}{2 }{\beta_{\rf}}}  \ket{p+m(v+v_0)}\!\bra{p+m(v+v_0)}\nn
&=&  \int \dd v\,\sqrt{\tfrac{m\beta'}{2\pi }}
\e^{-\frac{m v^2}{2}\beta'}\ketbra{p'\!+\!mv}{p'\!+\!mv} \nn
&=& 
\frac{1}{\sqrt{2 m\pi K_B T'}} \: \int \!\dd q\: \e^{-\frac{\left(q-p'\right)^2}{2m K_B T'}} \: \ketbra{q}{q}
\eea

We have produced the smearing \ $\rhocarr\mapsto \rhocarr'$ \ by a reference frame transformation, i.e.\ a passive transformation; \ we could have equally well used \ an active transformation.

A smearing becomes necessary of we consider either (or both) reference frames to be \emph{quantum} objects. Then all proper relative state (whether pure or not) will be characterized by finite, nonzero quantum uncertainties   $\Delta a$,  $\Delta v= \Delta p_{\rf}/M$, resp.\ of the relative displacement and velocity, fulfilling the Heisenberg relation
\be
 \Delta a\: \Delta v\: \ge\: \frac{\hbar}{2M}.
\ee
Since  $\Delta v>0$, also the  uncertainty  $\Delta E_{\rf}$ of the relative kinetic energy 
will be positive; together with the uncertainty $\Delta t_{\rf}$ of the relative time displacement
$ t_{\rf}$ of the two reference frames, they will have to satisfy the generalised energy/time uncertainty 
\be
 \Delta E_{\rf} \: \Delta t_{\rf} \:\geq\: \frac\hbar2
 \label{Etuncer}
\ee
(which, differently from the usual relation involving position and momentum,  is an effective relation, since there can be no self-ajoint time operator canonically conjugate to a Hamiltonian bounded from below).
Simultaneous measurements of both \ $E_{\rf} ,  t_{\rf}$ \ with  zero energy uncertainty are impossible; the best we can do is to smear the measurement, along the lines of what we described earlier. As usual for uncertain states, Gaussian smearing may minimise uncertainties. For a quantum  thermal distribution at temperature $T_{\rf}$ of $\RF$ wrt $\RF'$ one finds  again 
 $\Delta E_{\rf}= 1/{\beta_{\rf}}\sqrt{2}$. Replacing this in~\eqref{Etuncer} we obtain
\be
 \Delta t_{\rf} \:\geq\: \frac{\hbar}{\sqrt{2}\,K_BT_{\rf}} \:  .
 \label{Ttuncer}
\ee
This uncertainty will affect also the description of the state of ${\sf S}$ and of its evolution with respect to  $\RF'$, which involves the time $t'=t-t_{\rf}$  of the latter.

What we have found suggests relations among time, energy and temperature in a novel way, which does not use the usual techniques of going to a strip in the complex plane, or other similar techniques. We cannot elaborate further these hints without a more specific theory of QRFs. In some theories \cite{DopFreRob95,FioreLizziinpreparation} optimally localized states of $\RF$ with respect to  $\RF'$ are coherent states that can be represented by Gaussians; therefore there will be a formal analogy between the Gaussian of such pure states and that of the classical thermal state $\rhogru_{\beta_{\rf}}$ used above,
what might lead to relations analogous to~\eqref{Etuncer} and~\eqref{Ttuncer}.

\section{Final Remarks}

In this work, we have emphasized that the quantum nature of reference‐frame transformations must be treated on the same footing as the quantum nature of systems and frames themselves. By focusing on the simplest nontrivial symmetry—spatial translations—we have shown that even when a system $\mathsf S$ appears to be in a pure state relative to a classical frame $\RF$, allowing the translation “operator” (i.e., the frame transformation) to be nonpure generically produces a mixed state for $\mathsf S$ relative to another RF. This observation highlights a key conceptual point: classical reference frames, when described fully quantum mechanically, do not necessarily implement deterministic (pure) transformations between observers.
We have then illustrated the same phenomenon in the (slightly less simple) case of the symmetry group  of Galilei transformations among classical RFs for nonrelativistic quantum mechanics in 1+1 dimensions.
{In particular, we have shown that if the state of  frame $\RF$ w.r.t.\  frame $\RF'$ is thermal at some temperature $T_{\rf}$, then a quantum particle in the pure (improper) state with momentum $p$ (a zero temperature state)  w.r.t.\ $\RF$ appears in a thermal state with temperature $T'=T_{\rf} m/M$  w.r.t.\   $\RF'$. This can be considered as an instance of the phenomenon of QRF-relative thermodynamic equilibrium described in
section~7 of~\cite{HoehnEtAl:2023}. We have also sketched a possible link between thermal noise and the uncertainty in time translations between reference frames; in particular, how a (Boltzmann) thermal state of a reference frame w.r.t.\ the other (treated as a mixed Galilei boost) may  affect the energy/time uncertainty.}

Looking ahead, several directions suggest themselves immediately. First, a comprehensive treatment of more elaborate symmetry groups—including full Galilei transformations in higher dimensions, Poincaré transformations, and even discrete symmetries—will further clarify how generic quantum uncertainties in frames propagate into relational observables. Second, our companion work on quantum reference‐frame transformations in a noncommutative spacetime~\cite{FioreLizziinpreparation} will show how to extend these ideas to situations where the “arena” of events is itself quantum. 

Ultimately, a fully self‐consistent theory of quantum reference frames should address two aspects. On the one side a the possibility of mixed (and pure) frame transformations for arbitrary symmetry groups (or generalizations thereof), on the other, an  analysis of how those transformations act on arbitrary quantum systems. By recognizing transformations themselves as legitimate quantum objects we may gain a deeper understanding of the structure of quantum theories.

\subsection*{Acknowledgments}
We thank an anonymous referee for bringing references~\cite{TolouiGour} and~\cite{Meyer} to our attention, for comments on the proof that nonpure transformations are a semigroup, and mainly for  comments on the connections with thermal states.
We also thank Patrizia Vitale for discussions. 
We acknowledge support from the INFN Iniziativa Specifica GeoSymQFT. F.L.~acknowledges support from Grants No. PID2019–105614 GB-C21 and No. 2017-SGR-929, support by ICSC - Centro Nazionale di Ricerca in High Performance Computing, Big Data
and Quantum Computing, funded by European Union — NextGenerationEU. We also acknowledge the \textsl{Cost} actions CA21109: CaLISTA, CA23130: BridgeQG and CA23115: Relativistic Quantum Information.

\providecommand{\href}[2]{#2}\begingroup\raggedright\endgroup


\end{document}